\documentclass[print, twocolumn, showkeys]{revtex4-2}

\usepackage{lineno}
\usepackage[bf]{subfigure}
\usepackage{adjustbox}
\usepackage{amsmath}  
\usepackage{amsfonts} 
\usepackage{graphicx} 
\usepackage{multirow}
\usepackage{gensymb}
\usepackage{textcomp}
\usepackage{xcolor}
\usepackage{setspace}
\usepackage{soul}
\usepackage{cancel}

\usepackage{floatrow}
\usepackage{feynmp}
\RequirePackage[colorlinks,citecolor=blue,urlcolor=blue,linkcolor=blue]{hyperref}
\RequirePackage{url}

\definecolor{Red}{rgb}{1,0.0,0.0}
    \setulcolor{Red}
    
\begin{document}

\title{Search for the limits on anomalous neutral triple gauge couplings via $ZZ$ production in the $\ell \ell \nu \nu$ channel at FCC-hh}

\author{A.~Yilmaz} \email{aliyilmaz@giresun.edu.tr}
\affiliation{Department of Electrical and Electronics Engineering, Giresun University, 28200, Giresun, Turkey}

\begin{abstract}

This paper presents the projections on the anomalous neutral triple gauge couplings ($aNTGC$) via  $pp \rightarrow ZZ$  production in the 2$\ell$2$\nu$ final state at a 100 TeV proton-proton collider,  \verb"FCC-hh". 
The realistic \verb"FCC-hh" detector environments and its effects taken into account in the analysis. The study is carried out in the mode where one Z boson decays into a pair of same-flavor, opposite-sign leptons (electrons or muons) and the other one decays to the two neutrinos. The new bounds on the charge-parity (CP)-conserving couplings $C_{\widetilde{B}W} / \Lambda^{4}$ and CP-violating couplings $C_{WW}  / \Lambda^{4}$, $C_{BW} / \Lambda^{4}$ and $C_{BB} / \Lambda^{4}$ achived at 95\% Confidence Level (C.L.) using the transverse momentum of the dilepton system ($p_{T}^{\ell \ell}$) are $[-\, 0.042, \,\, +\,0.042]$,  $[-\,0.050, \,\, +\,0.050]$, $[-\,0.050, \,\, +\,0.050]$,  and $[-\,0.048, \,\, +\,0.048]$ in units of TeV$^{-4}$, respectively.


\end{abstract}
\maketitle

\section{Introduction}

Production of gauge boson pairs has an important role in the tests of the non-Abelian SU (2)$_{L} \times U (1)_{Y}$  gauge group of the electroweak sector of the Standard Model (SM) and exploring the new physics at the TeV scale.  Additionaly, diboson production is also connected to the spontaneous breaking of the EW gauge symmetry~\cite{Pich:2012sx, Neubauer_2011}.
For this reason,  a signature of new physics beyond SM may reveal themselves via the possible deviation from SM expected values in neutral triple tuning (NTG) couplings (which includes $ZZ\gamma$,  $Z\gamma \gamma$ and $ZZZ$ vertices).

New physics effects at high energy physics can be described in the terms of the Effective Field Theory (EFT) parameters. 
This parameters is comprehensive to indicate the most likely places to observe these effects; besides they cover the gauge symmetries of the SM and can be used at both tree level and loop level. 
In the framework of EFT theory, one can include the Anomalous NTG vertices in an effective Lagrangian and it can be described by CP-conserving and CP-violating couplings, even though there is no  electroweak NTGC exists at the tree level~\cite{PhysRevD.62.113011,Green:2016trm}. 
 
The dimension-eight (\texttt{dim-8}) effective Lagrangian for nTGC in the EFT framework  taking into account the local U(1)$_{EM}$ and Lorentz symmetry can be given as~\cite{Degrande_2014}

\begin{equation}
\mathcal{L}^{nTGC} = \mathcal{L}_{SM} + \sum_{i} \frac{C_{i}}{\Lambda^{4}} (\mathcal{O}_{i} + \mathcal{O}_{i}^{\dagger})
\label{eqn:lagrangian}
\end{equation}

where $i$ is defined as an index of equations working over the operators written as
\begin{widetext}
\begin{eqnarray}
\mathcal{O}_{\widetilde{B}W}  & = & i H^{\dagger} \widetilde{B}_{\mu\nu} W^{\mu \rho} \{D_{\rho}, D^{\nu} \} H,  \quad
\mathcal{O}_{BW}   =  i H^{\dagger} B_{\mu\nu}W^{\mu \rho} \{D_{\rho}, D^{\nu} \} H, \nonumber \\
\mathcal{O}_{WW}  & = & i H^{\dagger} W_{\mu\nu} W^{\mu \rho} \{D_{\rho}, D^{\nu} \} H, \quad
\mathcal{O}_{BB}   =  i H^{\dagger} B_{\mu\nu} B^{\mu \rho} \{D_{\rho}, D^{\nu} \} H.
\label{eqn:lagrangian2}
\end{eqnarray}
\end{widetext}

where $\widetilde{B}_{\mu \nu}$ is the dual $B$ strength tensor. The layout given in the formula of the operators
\begin{widetext}
\begin{eqnarray}
B_{\mu\nu} & = & (\partial_{\mu} B_{\nu} - \partial_{\nu} B_{\mu}),  \quad  
W_{\mu\nu}  =  \sigma^{I} (\partial_{\mu} W^{I}_{\nu} - \partial_{\nu} W^{I}_{\mu} + g\epsilon_{IJK} W^{J}_{\mu} W^{K}_{\nu} )
\label{eqn:lagrangian3}
\end{eqnarray}
\end{widetext}

with $\langle \sigma^{I} \sigma^{J} \rangle$ = $\delta^{I \, J} / 2$ and $D_{\mu} \equiv \partial_{\mu} - i g_{w} W^{i}_{\mu} \sigma^{i} - i \frac{g^{\prime}}{2} B_{\mu} Y $.

It is expected that the main contribution of new physics to the amplitude of $ff \to ZZ$ process is coming from the interference between the SM and the \texttt{dim-8} operators when the new physics energy scale is high.
The square of the amplitude with \texttt{dim-8} operators does not carry a significant contribution from the heavy new physics except that the interferences between the SM and the \texttt{dim-8} and dimension-ten (\texttt{dim-10}) operators are excessively suppressed.

The \texttt{dim-6} operators can influence on nTGC at one-loop level (at the order $\mathcal{O}$($\alpha \hat{s} / 4\pi  \Lambda^2$) despite being not effective at the tree level~\cite{Degrande_2014}. 
Nevertheless, the tree level contributions from \texttt{dim-8} operators are of the order $\mathcal{O}$($\hat{s}  v^{2} / \Lambda^4$).
For this reason, a one-loop contribution of the \texttt{dim-6} operators can be neglected in accordance with \texttt{dim-8} operators considering $\Lambda \lesssim 2v \sqrt{\pi / \alpha}$.

The coefficients of  four \texttt{dim-8} operators describing in Eq.~\ref{eqn:lagrangian2} are the CP-conserving couplings $C_{\widetilde{B}W} / \Lambda^{4}$ and  CP-violating couplings  $C_{WW}  / \Lambda^{4}$, $C_{BW} /   \Lambda^{4}$ and $C_{BB}  /  \Lambda^{4}$. 
These are couplings of the \texttt{dim-8} operators transformed from the aTGC given in the Ref.~\cite{Degrande_2014} 

The $\ell \ell \nu \nu$ final state via the production of  $ZZ$ dibosons  has been carried out  at Fermilab (CDF) collaboration ~\cite{CDF:2011ab, PhysRevD.89.112001} and $D\O$ collaborations~\cite{Abazov:2008yf}. Recently, the ATLAS~\cite{Aaboud:2019lgy} ( at $\sqrt{s}= 13$ TeV, corresponding to an integrated luminosity of $36.1$ fb$^{-1}$) and CMS~\cite{Sirunyan:2017jtu} experiments also reported the bounds on the EFT parameters using the $\ell\ell\nu\nu$ final state.
 This high energy results in an improvement of the cross section, that expands the scope of triple gauge coupling studies. Additionally numerous phenomenological studies have been carried out the investigating the limits of aNTGCs at hadron colliders in the EFT framework~\cite{Senol_2014, Mangano_2016, Frye_2016, DORIGO2018211, SENOL2018365, Senol:2019qyl, KHANPOUR2020115141}. 
Table~\ref{tab:limits1} summarise the latest experimental limits on dim-8 couplings by LHC~\cite{Sirunyan:2020pub} $ZZ\rightarrow 4l$ proceses at $\sqrt{s}=13$ TeV with $L_{int} = 137$ fb$^{-1}$ and phenomenological limits by Ref.~\cite{Yilmaz_2020} on $C_{\widetilde{B}W} / \Lambda^{4}$, $C_{WW}  / \Lambda^{4}$, $C_{BW} /   \Lambda^{4}$ and $C_{BB}  /   \Lambda^{4}$ EFT parameters with the same final state using the 100 TeV FCC-hh option with $L_{int} = 10$ ab$^{-1}$. In this table, all couplings are set to zero except one that we are working on it.\\

\begin{table}[h!]
\centering 
\caption{Current observed one dimensional 95\% confidence level (C.L.) bounds on the coefficients of \texttt{dim-8} operators $C_{\widetilde{B}W} / \Lambda^{4}$, $C_{WW}  / \Lambda^{4}$, $C_{BW} /   \Lambda^{4}$ and $C_{BB}  /   \Lambda^{4}$ by using the transformation of the EFT parameters from Ref.~\cite{Degrande_2014}}

\begin{tabular}{  lccc }
 \hline 
 Couplings &  \multicolumn{3}{c}{Limit 95\% C.L.}    \\
  $ (TeV^{-4})$   &  & $ZZ \rightarrow 4 \ell$~\cite{Sirunyan:2020pub}  & $ZZ \rightarrow 4 \ell$~\cite{Yilmaz_2020}\\
\hline
$ C_{B W}   / \Lambda^{4}$    &      & $-\,2.3, \,\, +\,2.5$          & $-\,0.09, \,\, +\,0.09$\\
$ C_{BB} / \Lambda^{4}$        &     & $-\,1.4, \,\, +\,1.2$          & $-\,0.10, \,\, +\,0.10$ \\
$ C_{WW} / \Lambda^{4}$      &     & $-\,1.4, \,\, +\,1.3$          & $-\,0.21, \,\, +\,0.21$  \\
$ C_{\widetilde{B}W}  / \Lambda^{4}$  & & $-\,1.2, \,\, +\,1.2$  & $-\,0.26, \,\, +\,0.26$ \\
\hline 
\end{tabular} 
\label{tab:limits1}
\end{table}

For the future collider project, there are proposals for the future higher energy hadron colliders to carry out the researches directly at the energy frontier. These proposals include FCC-ee, FCC-eh and FCC-hh collider types working at different center of mass energies.
The hadron collider option of FCC (FCC-hh) is planned to run at the center of mass energy of 100 TeV and the integrated luminosity of  1 ab$^{-1}$(initial) and 30 ab$^{-1}$ (ultimate)~\cite{Abada_2019physOp, Abada_2019FCChh}. 

Searching the new physics effects in the production of a diboson is requiring great effort.
In the literature ZZ diboson production has been examined in two decay channels, such as the  ``4$\ell$"  and  ``$\ell\ell\nu\nu$" channel~\cite{Green:2016trm}.  
In the $ZZ \rightarrow$ 4$\ell$ decay channel,  both of the Z bosons decay into two same-flavor, oppositely charged leptons. This process gives rise to  the inclusion of a very low background, being kinematically reconstructable in the final state. 
In the $ZZ \rightarrow$ $\ell\ell\nu\nu$ channel,  one of the Z decays into a same-flavor, oppositely-charged two leptons, while the other one decays into neutrinos, which leads to an increase in the missing transverse energy in the final state. 
The branching ratio of ``$\ell\ell\nu\nu$"  final state is greater than the ``4$\ell$"  final state and the sensitivity to anomalous triple gauge couplings (aTGCs) would be higher than the ``4$\ell$"  final state.
``$\ell\ell\nu\nu$"  final state is still exposed to a larger background contamination, and it requires strict experimental selection which leads to force one Z boson boosted against the other in the transverse plane is necessary to retain the background at a more feasible level~\cite{PhysRevD.78.072002}.
Hence, the ``$\ell\ell\nu\nu$"  final state includes more data quantities than the ``4$\ell$" final state for the events with high-$p_T$ Z bosons, and this final state also presents competitive precision for integrated and differential measurements, as well as good sensitivity to aTGCs~\cite{Aaboud_2019}. 
Therefore, the $\ell\ell\nu\nu$ final state is chosen to analyze using the FCC-hh option could be open a new possibility to extent the most stringent upper limits on EFT parameters thanks to its high center-of mass energy 100 TeV and  capability to reach $\mathcal{L}_{int}= $ 30 ab$^{-1}$ (ultimately).

This paper is arranged as follows:  In Sect.~\ref{sec:simulation} provides details of the simulation environment for the $ZZ$ diboson signal and background production samples at the FCC-hh collider. Sect.~\ref{sec:eventSelect} describes the algorithms developed for event selection strategies of this phenomenological study in the $\ell\ell\nu\nu$ final state.
Obtained results for  the $\ell\ell\nu\nu$ final state analysis are given in Sect.~\ref{sec:results}. 
Finally,  conclusions on the main results and findings of each couplings are given in Sect.~\ref{sec:conclusion}.

\section{Generation of signal and background events} \label{sec:simulation}
\begin{figure}[!htb] 
 	\centering 
	\subfigure[]{%
 		  \includegraphics[width=1\textwidth]{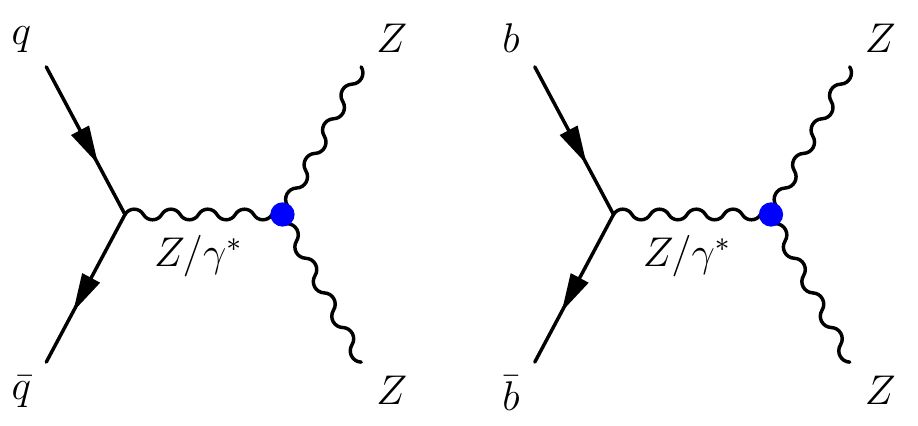}
		\label{fig:sig}
    	}\\
    	\subfigure[]{%
   		   \includegraphics[width=1\textwidth]{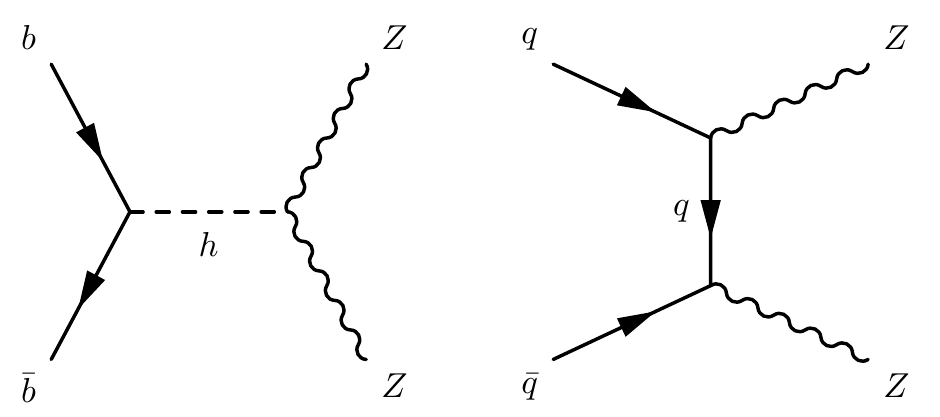}
		\label{fig:bck}
    	}
  \caption{Representative Feynman diagrams of ZZ process with an aNTGC vertex (blue dot) \textbf{(a)} for signal and \textbf{(b)} SM background} 
   \label{fig:feynmanDiag}
\end{figure}

In this section, we present the production of the $pp \rightarrow ZZ \rightarrow  \ell^+ \ell^- \nu \bar{\nu} $ signal events as well as the background events (SM) taking into account the experimental conditions of FCC-hh.  In order to find the bounds on the  \verb"aNTG"  couplings, we used  the  \verb"UFO" model file~\cite{Degrande_2012}  implemented into Monte Carlo (MC) event generator \verb"MadGraph5_aMC@NLO v2.6.4"~\cite{Alwall_2014}.
The \verb"PYTHIA v8.2"~\cite{SJOSTRAND2015159} package is utilized for the parton showering, fragmentation and hadronization of generated signal and background events.
The \verb"LHAPDF v6.1.6"~\cite{Buckley_2015} library with \verb"NNPDF v2.3"~\cite{BALL2013244, Ball_2017} set is a default set of parton distribution functions (PDFs) used to produce all MC samples.  The jets were collected by using \verb"FastJet"~\cite{Cacciari:2011ma} where the anti-k$_T$ algorithm~\cite{Cacciari:2008gp} already implemented with a cone radius is R = 0.4.  In the search of new physics effects, $3 \times 10^{6}$ events are generated for the signal process, $pp\rightarrow ZZ$, as well as the background process  (in the same final state with signal process) by varying one coupling value at a time on each of the \texttt{dim-8} couplings.
The realistic detector effects are taken into account by using the FCC-hh detector card presents inside the \verb"Delphes v3.4.1"~\cite{de_Favereau_2014}. All generated events are analyzed by using the \verb"ExRootAnalysis"~\cite{exRootAnalysis} package  with \verb"ROOT v6.16"~\cite{BRUN199781}. The expected event numbers given in the plots of kinematic variables are weighted with the cross-section of each process including the branching times integrated luminosity of $\mathcal{L}_{int}$ = 10 ab$^{-1}$.
Since Feynman diagrams may simplify and vizualize the aNTGC vertices clearly, the leading order Feynman diagrams contibuting to the signal and background processes are depicted in Fig.~\ref{fig:sig} and Fig.~\ref{fig:bck}, respectively. The blue dot indicates the aNTGC vertex in the $ZZ$ process where one of Z decays into $\ell\ell$ and the other one decays into $\nu\nu$ final states.

\begin{figure} [hbt]
 \centering
   \includegraphics[width=1\textwidth]{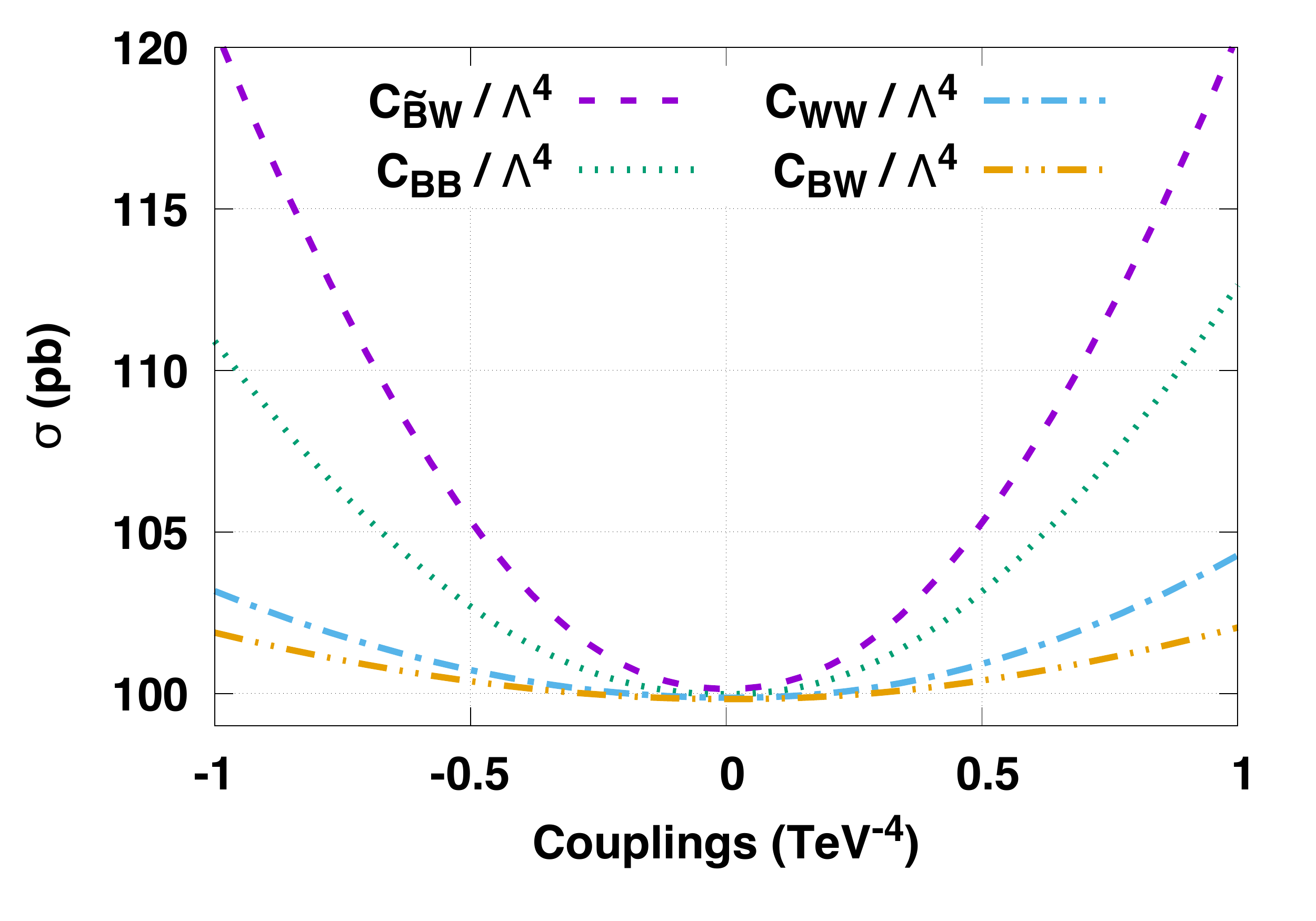}
  \caption{The total cross sections as a function of CP-conserving and CP-violating coupling terms in the Lagrangian at FCC-hh} 
   \label{fig:xSectionPlot}
\end{figure}

In Fig.~\ref{fig:xSectionPlot}, we depicted the cross-sections of the $ZZ$ process in unit of [pb]  as a function of the energy [TeV$^{-4}$] for the  \texttt{dim-8} couplings at the generator level where the default mass of the $Z$ boson is chosen as 91.187 GeV~\cite{pdg2018}.  This plot is produced by changing only one coupling value at a time in the range of  $\pm 1$  [TeV$^{-4}$] while the others are set to zero.

The deviation from the SM value shows that the main contribution for the signal is seen in the CP-conserving coupling (see Fig.~\ref{fig:xSectionPlot}).  So the effective \texttt{dim-8} aNTG operators and a SM contribution as well as interference between effective operators and SM contributions are considered  in the analysis.


\section{Event selection} \label{sec:eventSelect}

This analysis considers the $pp \rightarrow ZZ $ in the $\ell\ell\nu\nu$ final state based on Ref.~\cite{Aaboud_2019}. 
The candidate event selection procedure against larger background is optimized to handle with the contaminations.
In this topology the two Z bosons originating from the aNTGC desired to be in opposite direction (back-to-back). 

The pre-selection requirement in the analysis is the existence of one dilepton of the same flavor with opposite charge to construct the Z-boson.
The missing transverse momentum ${\vec{E}}^{miss}_{T}$ is computed as a negative vector sum of all charged leptons and jets in the event.  
The events are required to have at least 2 leptons ($N_{\ell} \ge 2$) of the same flavor with opposite charge ($e^{+} e^{-}$ or $\mu^{+} \mu^{-}$). Transverse momentum of leading lepton $\ell^1$, (subleading lepton $\ell^2$), $p_{T} > 30 \, (20)$ GeV is required and the cut on the pseudo-rapidity between two leptons is also applied as $|\eta^{\ell} |<$ 2.5. 
Events with relatively a few calorimeter activity are rejected by vetoing on the presence of more than one jet with $p_{T} > 20$ and $|\eta^{j} | < 4.5$ in the detector. 
The events selection is optimized by imposing the transverse momentum balance ratio $(p_T^{miss} - p_T^{\ell\ell}) / p_T^{\ell\ell} < 0.3 $. 
The distance $\Delta R $ between two objects in $\eta$-$\phi$ plane is evaluated by the function $\Delta R = \sqrt{\Delta \eta^{2}  + \Delta \phi^{2}}$
and we accepted the leptons only if the distance between lepton and jet, $\Delta R_{\ell j} $ is greater than 0.4 for further reducing the contributions of overlapping jet. 
In addition to previous cuts 
requiring the events having any extra lepton with $p_{T} > 10$ GeV. 
Suppressing the effect of jet energy scale uncertainties, we applied extra cuts on the jets which are selected to have $p_{T} > 35 $ GeV for the central region$|\eta| < 2.4$ and  $p_{T} > 40 $ GeV for the forward region $2.4 < |\eta| < 4.5$. The dilepton invariant mass.($m_{\ell\ell}$) is required to be within 15 GeV of the nominal Z boson mass. 
Candidate events are required to have $\cancel{E}_{T}^{miss} > 110$ GeV, 
and the distance $\Delta R_{\ell \ell} $ between two leptons in $\eta$-$\phi$ plane is required to be greater than 2.1 which imposes the leptons must be close to each other.  Only if the azimuthal angle difference between the missing transverse momentum $\vec{\cancel{E}}_T^{miss}$ and the dilepton system, $\Delta\phi (\vec{\cancel{E}}^{miss}_{T}, \vec{p}_{T}^{\, \ell \ell}) > 2.5$ radian is accepted. 
Finally, the candidate events constructed from $p^{\ell\ell} $ is used for further analysis.

\begin{figure} [hbt]
 \centering
 \subfigure[]{
   \includegraphics[width=0.45\textwidth]{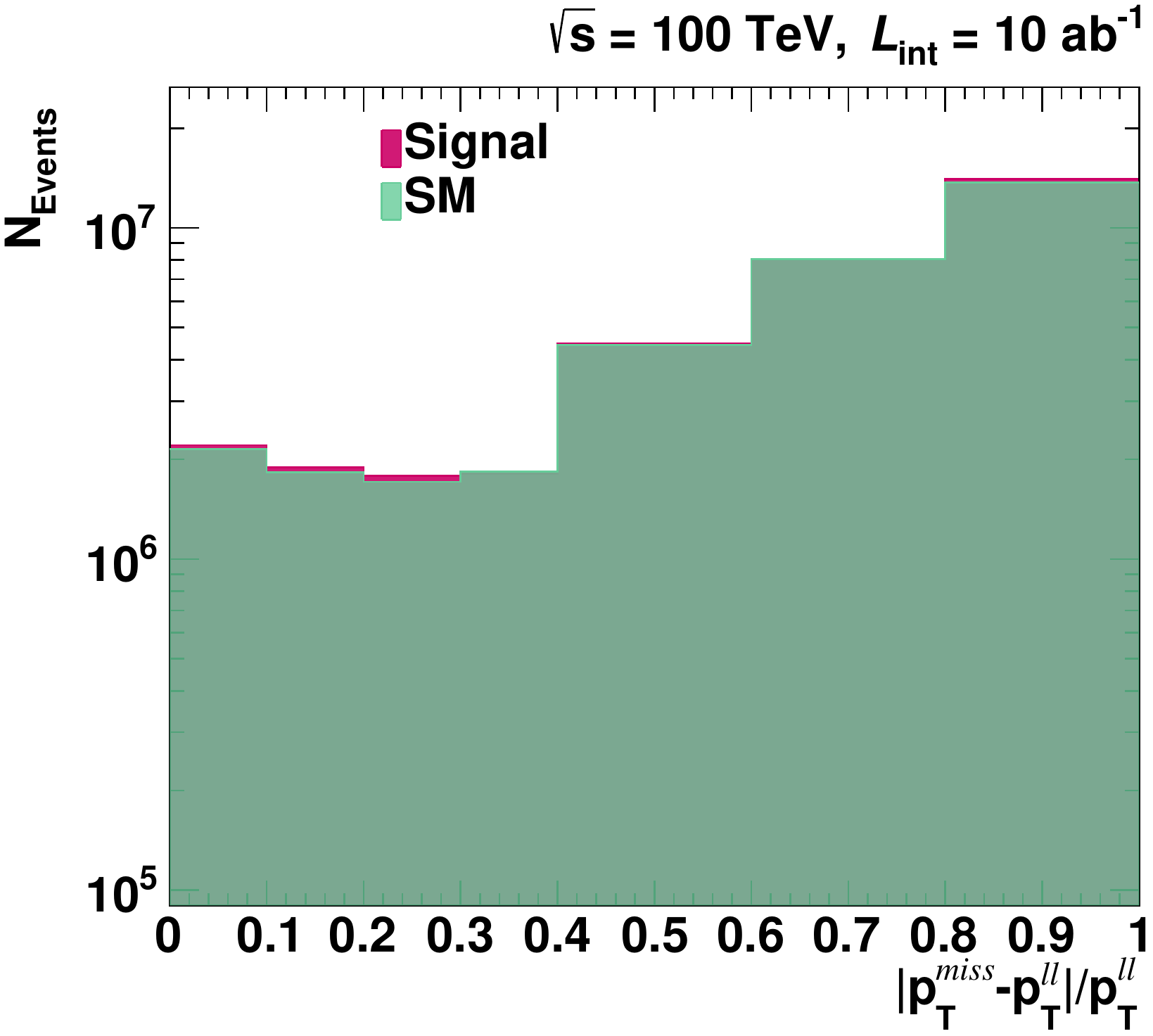} \label{fig:pTbalanceR}
   }
   \subfigure[]{
   \includegraphics[width=0.45\textwidth]{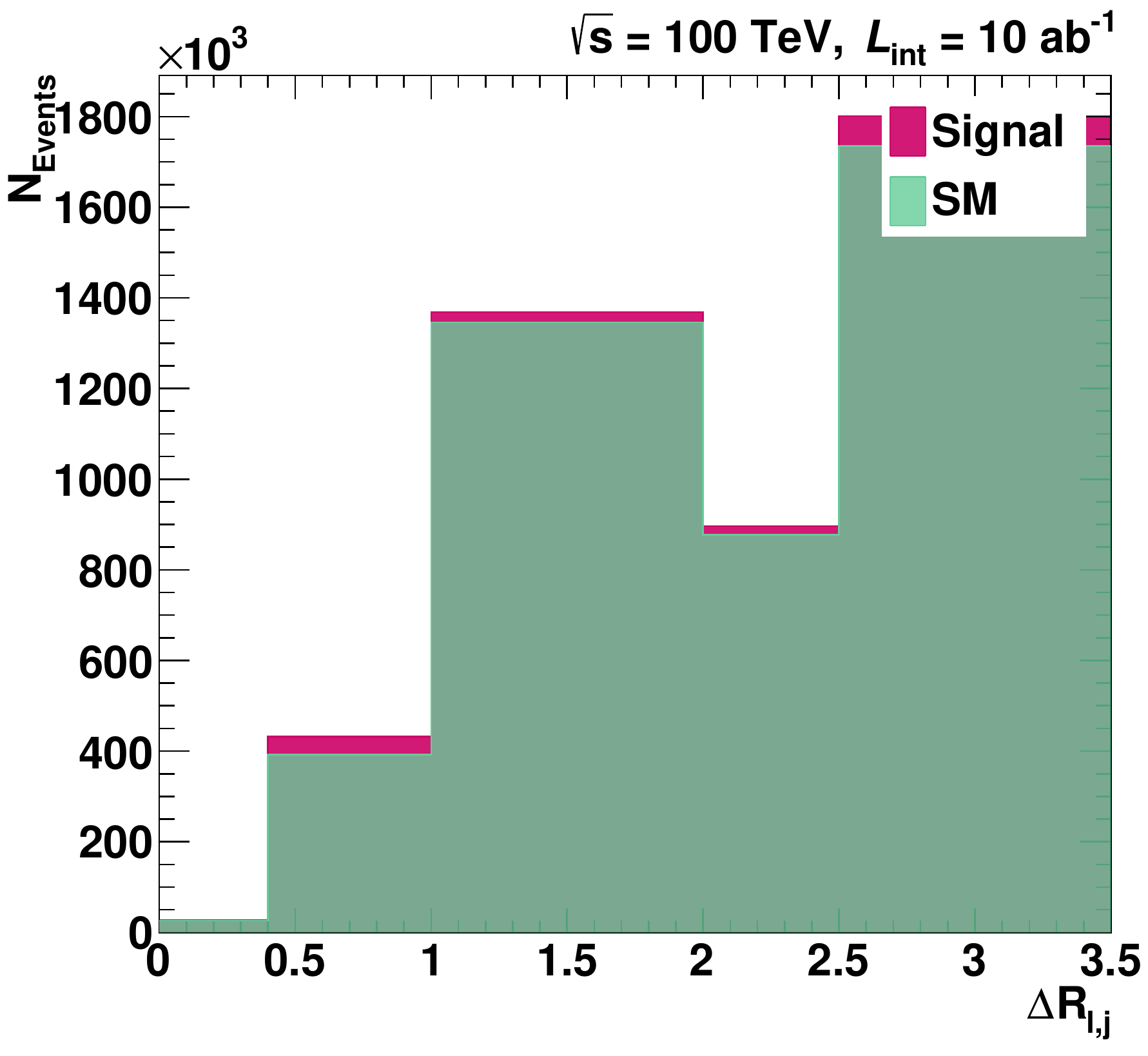} \label{fig:deltaPhil}
   }
   \subfigure[]{
   \includegraphics[width=0.45\textwidth]{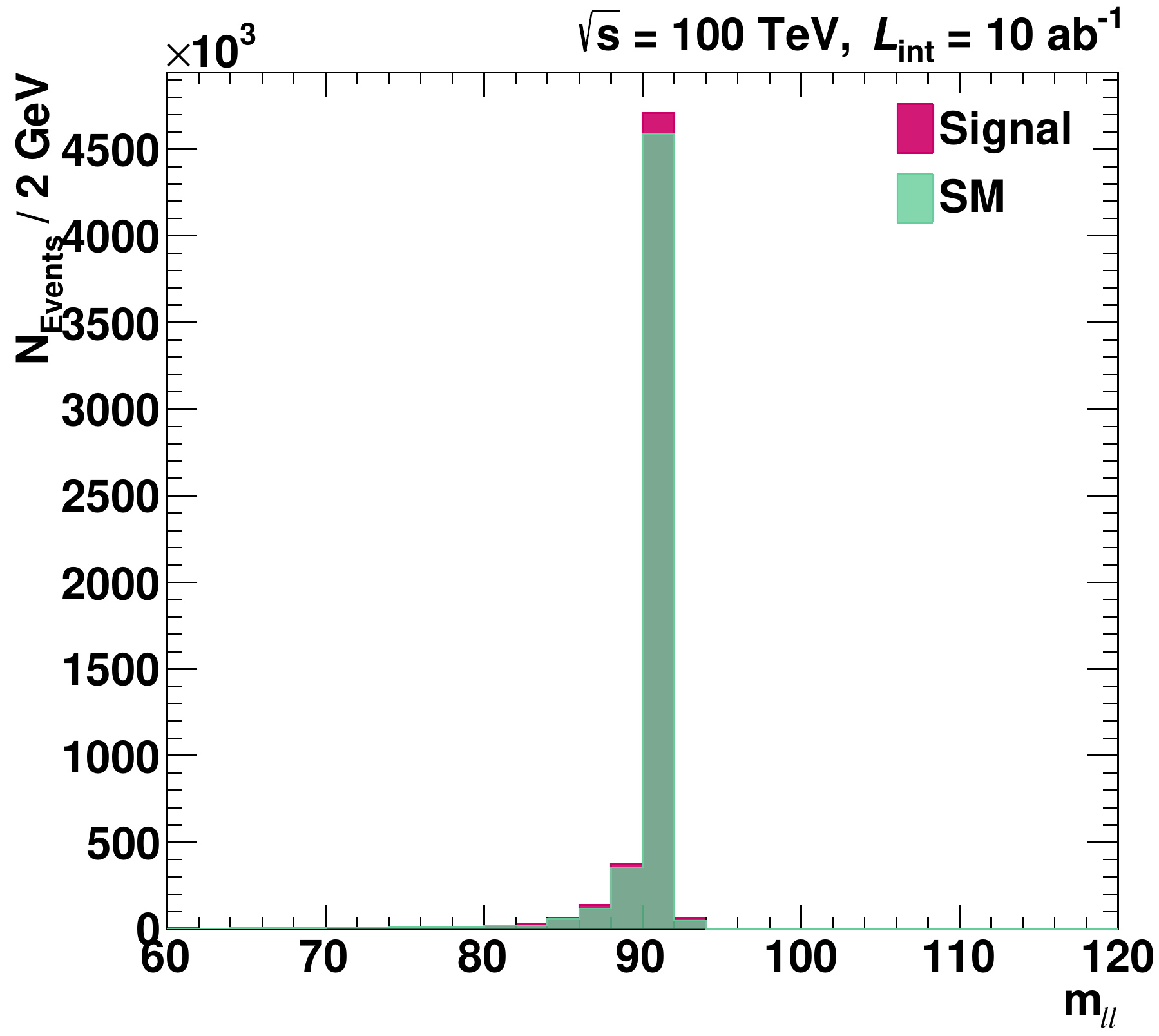} \label{fig:bJets}
   }
   \subfigure[]{
   \includegraphics[width=0.45\textwidth]{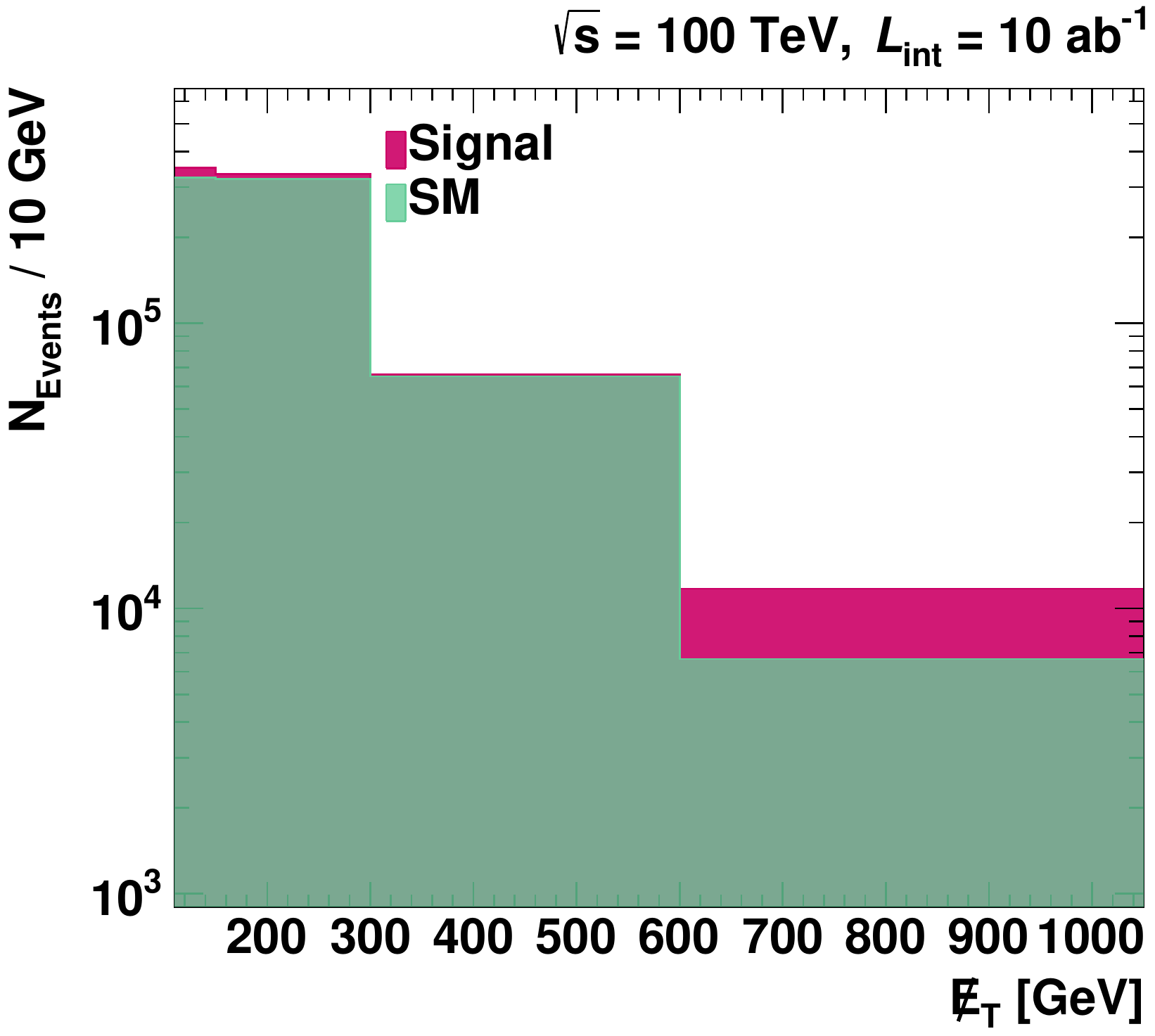} \label{fig:metCut}
   }
   \caption{\label{fig:kinPlots1} \textbf{(a)} The distribution of transverse momentum balance ratio $|(\cancel{E}^{miss}_{T} - p_{T}^{\, \ell \ell})|  / p_{T}^{\, \ell \ell}$, \textbf{(b)} the distance between lepton and jet $\Delta R (\ell, j)$, \textbf{(c)} the reconstructed invariant mass distribution of dilepton system and \textbf{(d)} distribution of the reconstructed $\cancel{E}_{T}^{miss}$ for the bins with $\cancel{E}_{T}^{miss} > $110 GeV. The signals are compared with the SM background which are most probable aTGC contributions for different values of the EFT parameters. The results are shown for  CP-conserving coupling, $C_{\widetilde{B}W}  / \Lambda^{4}$ = 5 TeV$^{-4}$ at $\mathcal{L}_{int}=$ 10 ab$^{-1}$. Each objects includes the implementation of all the cuts in Table~\ref{tab:cutTable}  prior to the cut on that variable} 
\end{figure}

\begin{figure} [hbt]
 \centering
    \subfigure[]{
   \includegraphics[width=0.45\textwidth]{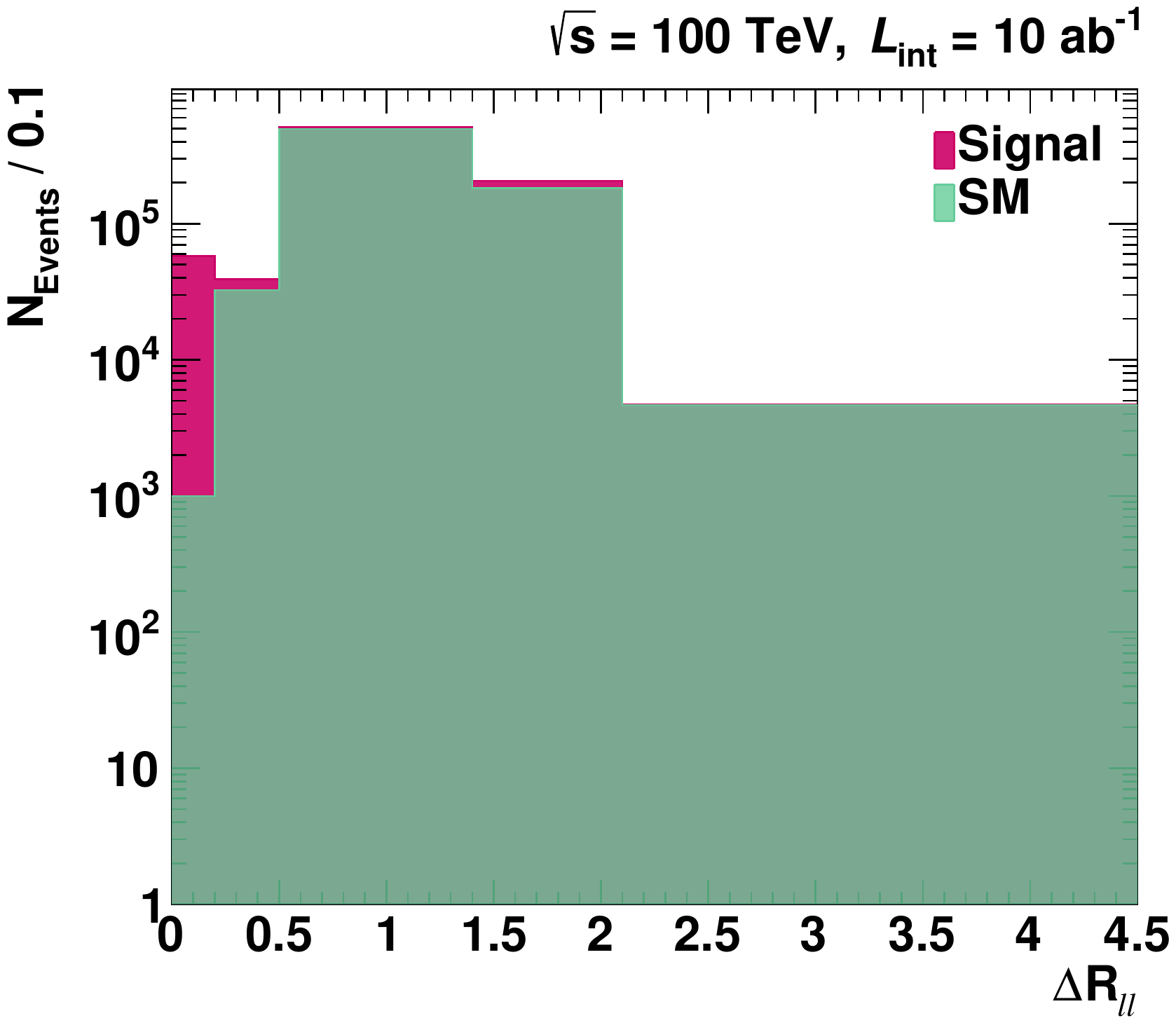} \label{fig:bJets}
   }
   \subfigure[]{
   \includegraphics[width=0.45\textwidth]{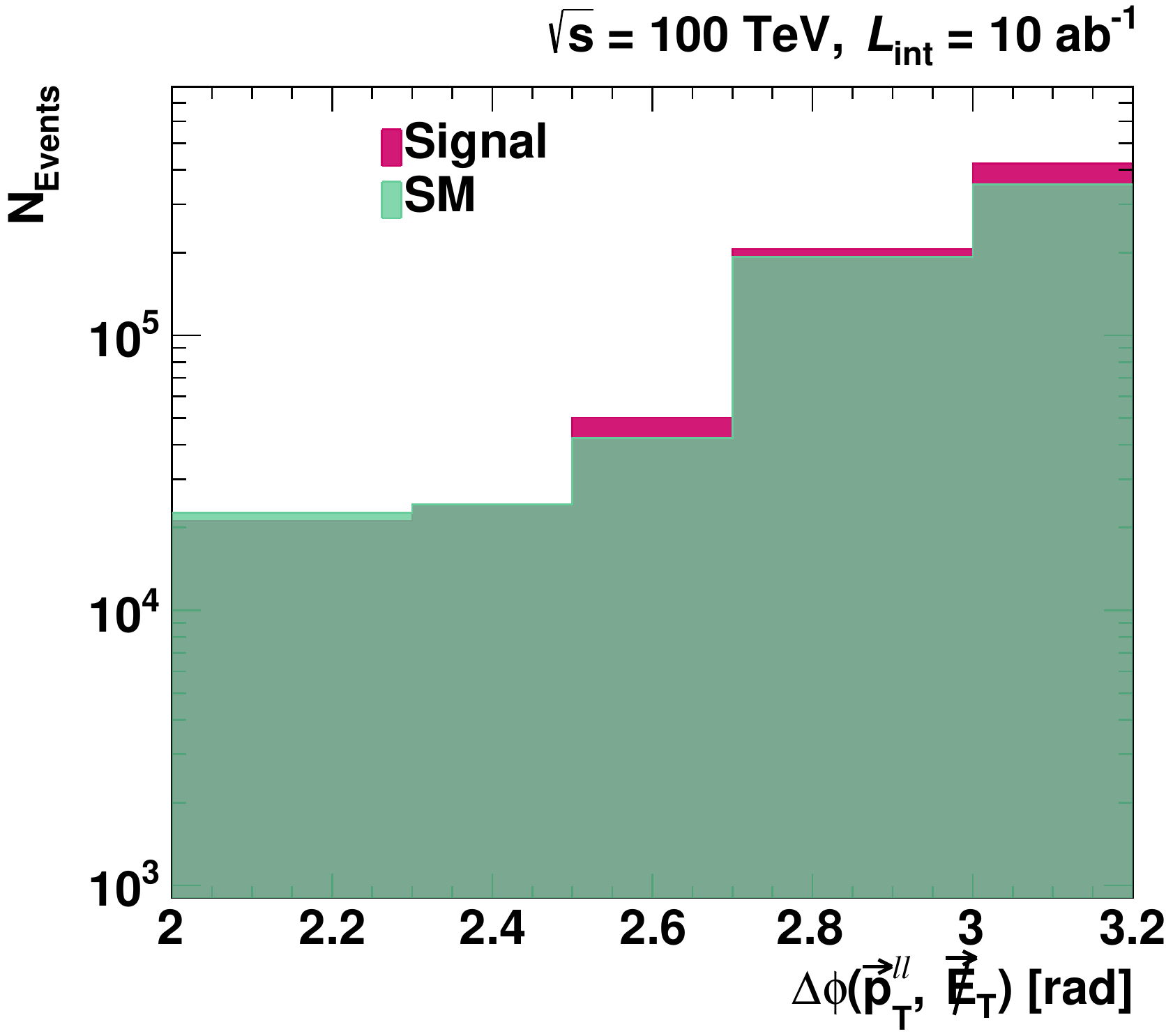} \label{fig:bJets}
   }
  \caption{\textbf{(a)} The distance $\Delta R_{\ell \ell}$ between two leptons in $\eta$-$\phi$ plane and \textbf{(b)} the distribution of the azimuthal angle between missing transever momentum and dilepton system, $\Delta\phi (\vec{\cancel{E}}^{miss}_{T}, \vec{p}_{T}^{\, \ell \ell}$)  
  Each variable in the distributions respects the implemented cuts on that variable accordingly. The signals are compared with the SM prediction which are possible aTGC contributions for different values of the strength of the coupling parameters. The results are shown for CP-conserving coupling, $C_{\widetilde{B}W}  / \Lambda^{4}$ = 5 TeV$^{-4}$ at $\mathcal{L}_{int}=$ 10 ab$^{-1}$
  } 
   \label{fig:kinPlots2}
\end{figure}

The kinematical distributions used to probe the signal from background in this analysis are plotted in Fig.~\ref{fig:kinPlots1} and Fig.~\ref{fig:kinPlots2} where each plot is made with the implementation of all the cuts sequentially on that variable, according to the cut flow given in Table~\ref{tab:cutTable}.

The all steps of cut flow in the analysis for selecting the events are summarized in Table~\ref{tab:cutTable}.

\begin{table*}[!htp]
\caption{Definition of sequential cut selections applied in the analysis of signal and background
samples \label{tab:cutTable}}
\centering%
\resizebox{\hsize}{!}{
\begin{tabular}{lll}
\hline 
Cuts index &  & Object requirements \tabularnewline
\hline 
Cut-0  & & Preselection: $N_{\ell (e, \mu)}>=$ 2 and a pair of  leptons; opposite-sign with same-flavor lepton \tabularnewline
Cut-1  & & Transverse momentum of a dilepton pair and pseudo-rapidity:  \tabularnewline
	   &&  $p_{T}^{(\ell^{1}, \ell^{2})} >$ (30, 20) GeV, and $|\eta^{\ell}| < 2.5 $  \tabularnewline
Cut-2   & &Transverse momentum of Jets and  its pseudo-rapidity: \tabularnewline
	   & & $p_{T}^{j} >$ 20 GeV, and $|\eta^{j}| < 4.5 $  \tabularnewline 
Cut-3  &  &$p_T$ balance: $|(\cancel{E}^{miss}_{T} - p_{T}^{\, \ell \ell})|  / p_{T}^{\, \ell \ell} < 0.3 $   \tabularnewline 
Cut-4  &  &$\Delta R (\ell, j) >$ 0.4 between lepton and jet \tabularnewline
Cut-5  &  &Lepton rejection: reject any additional lepton with $p_{T} > 10$ GeV  \tabularnewline
Cut-6  &  &Invariant mass: $M_{inv}^{rec}(\ell \ell)< 15 $ GeV   \tabularnewline
Cut-7  &  &Hard Jets: $p_{T}^{j} > 35$ GeV for $ |\eta^{j}| < 2.4$ and $p_{T}^{j} > 40$ GeV for  $2.4 < |\eta^{j}| < 4.5 $\tabularnewline
Cut-8  &  &$\cancel{E}^{miss}_{T} > 110 $ GeV   \tabularnewline
Cut-9  &  & $|\Delta R (\ell, \ell)| >$ 2.1 between leptons   \tabularnewline
Cut-10  & & $\Delta\phi (\vec{\cancel{E}}^{miss}_{T}, \vec{p}_{T}^{\, \ell \ell}) > 2.5$ radian   \tabularnewline
\hline 
\end{tabular}
}
\end{table*}

\begin{figure} [hbt]
 \centering
   \includegraphics[width=1\textwidth]{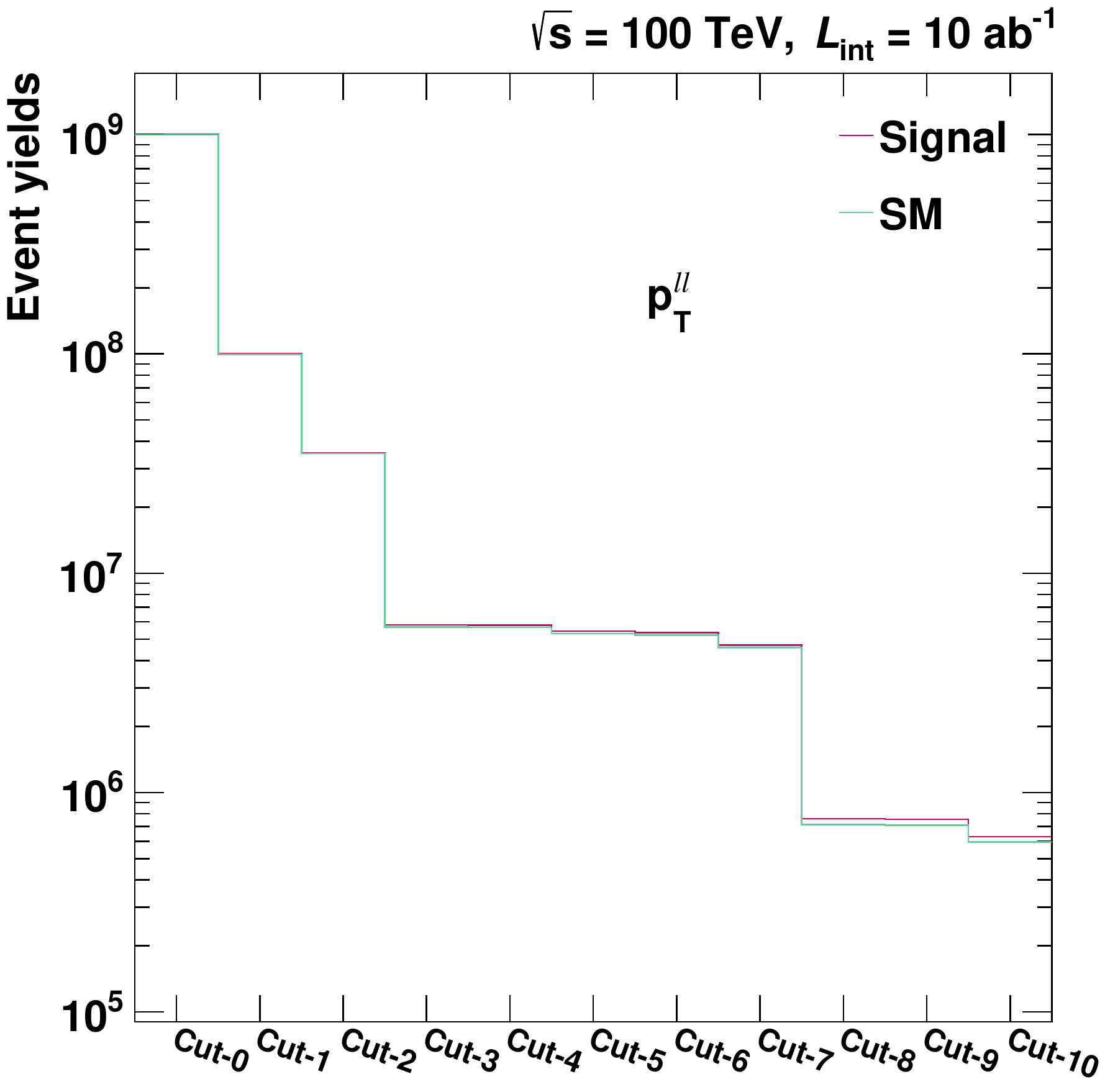}
  \caption{A weighted cutflow of the events selection through the $p_{T}^{\ell \ell}$ for each cut defined in the Table~\ref{tab:cutTable}} 
   \label{fig:cutEff}
\end{figure}

After applying the kinematical cuts discussed above,  the effects of each cut on the reconstructed transverse momentum, $p_T^{\ell\ell}$ and the obtained event yields is depicted in Fig.~\ref{fig:cutEff} for CP-conserving coupling, $C_{\widetilde{B}W}  / \Lambda^{4}$ = 5 TeV$^{-4}$ at $\mathcal{L}_{int}=$ 10 ab$^{-1}$,  as well as the other couplings have also similar behavior of the distributions. 

\section{Results}\label{sec:results}

\begin{figure}[!htb] 
 	\centering 
		\subfigure{ 
 		  \includegraphics[width=0.45\textwidth]{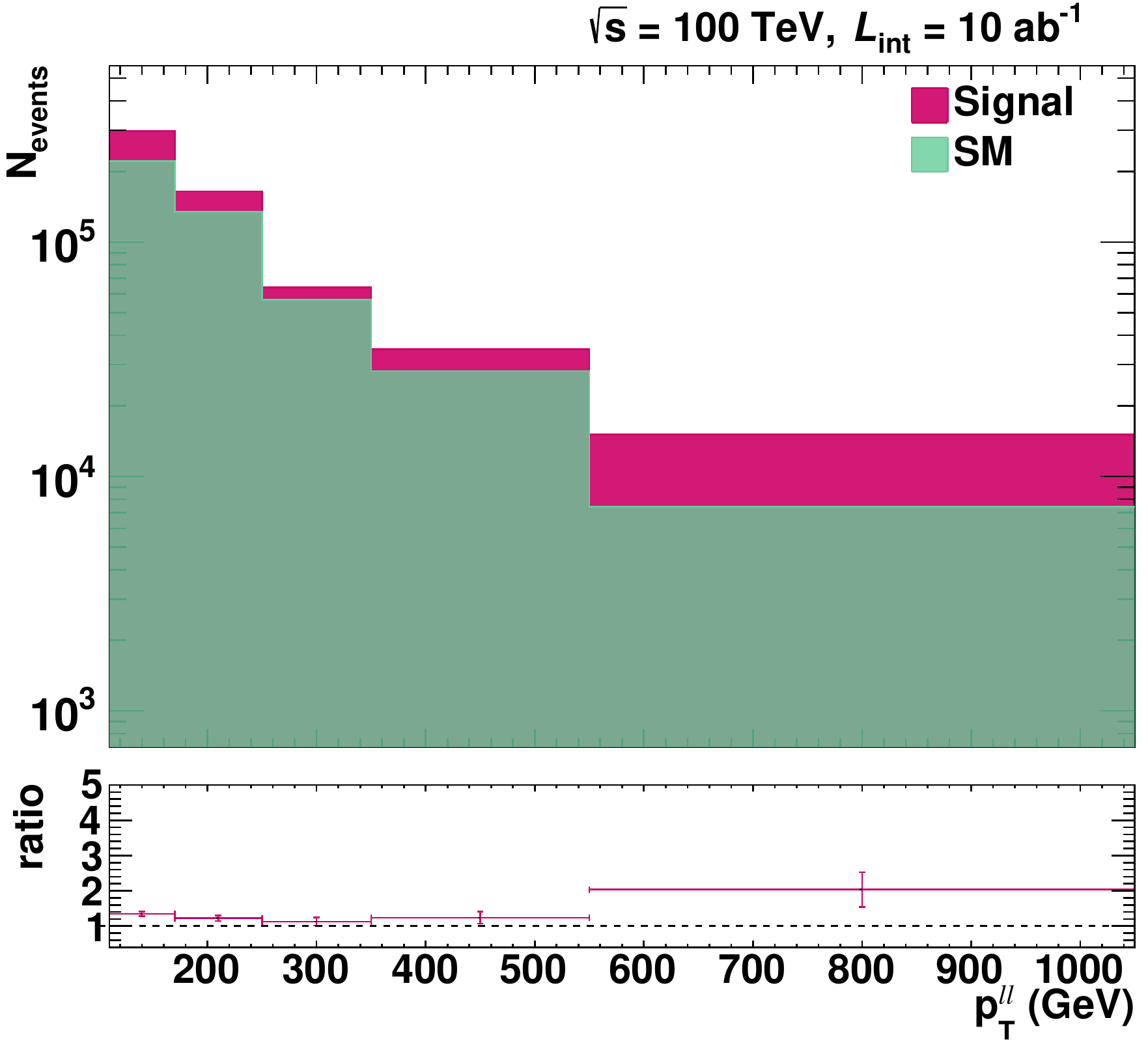}
		  }
		  \subfigure{ 
 		  \includegraphics[width=0.45\textwidth]{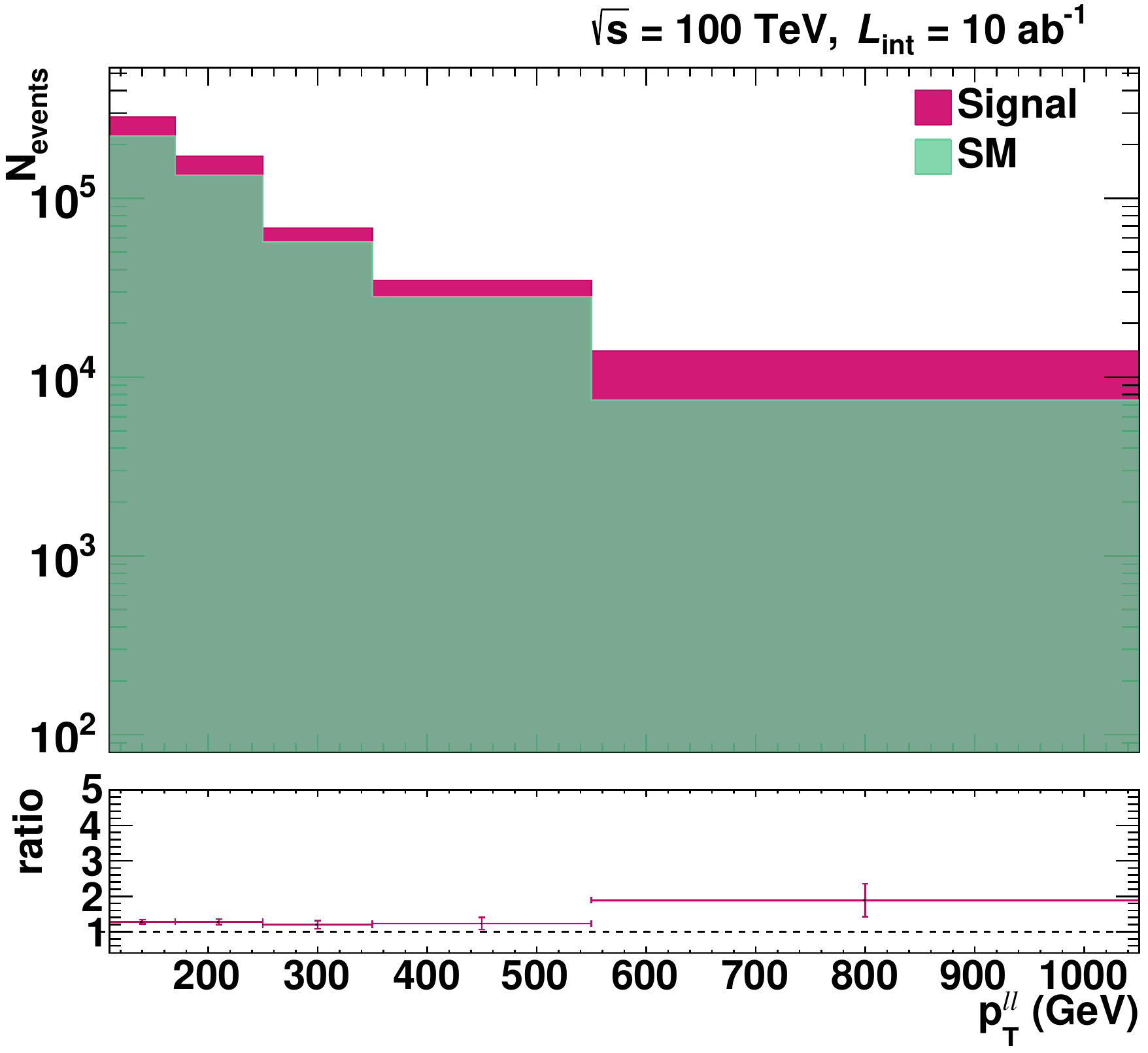}
		  }
		  \subfigure{ 
 		  \includegraphics[width=0.45\textwidth]{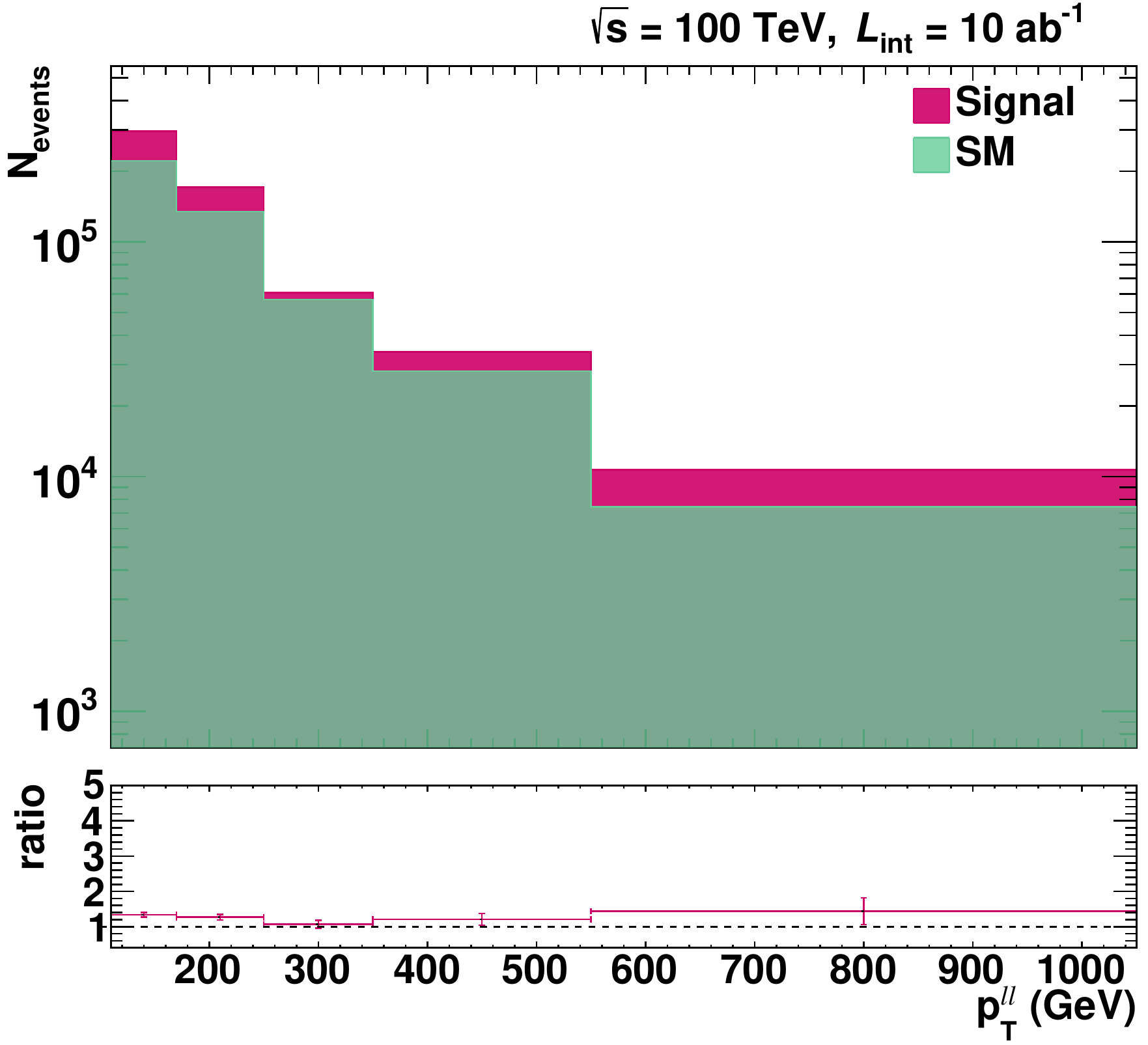}
		  }
		  \subfigure{ 
 		  \includegraphics[width=0.45\textwidth]{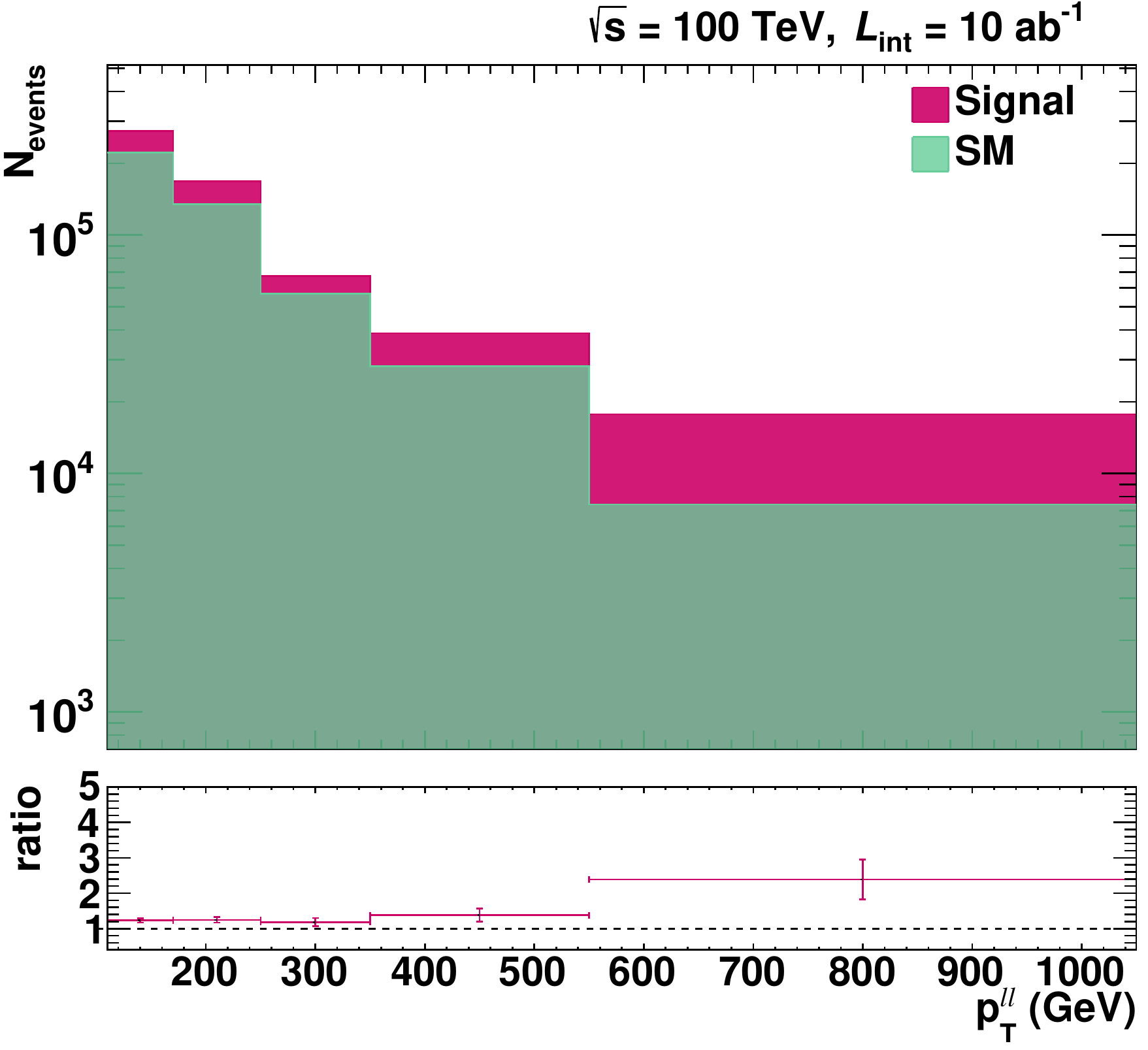}
		  }
     \caption{Distribution of the reconstructed $p_{T}^{\ell \ell}$ for the bins with $p_{T}^{\ell \ell} > $110 GeV.The signal samples are compared with the SM background that are potential aNTGC contributions for each values  of the EFT parameters . The results are shown individually for  $C_{\widetilde{B}W}  / \Lambda^{4}$ (top left), $C_{BW}  / \Lambda^{4}$ (top right), $C_{WW}  / \Lambda^{4}$ (bottom left), and  $C_{BB}  / \Lambda^{4}$(bottom right) which are set to 5 TeV$^{-4} with $$\mathcal{L}_{int}=$ 10 ab$^{-1}$. In the $p_{T}^{\ell \ell}$ distribution, bin contents are normalized to the bin widths. The lower plot shows the ratio of signal and background in the bins. The error bars in $x$-axis represent the variable bin size in the $p_{T}^{\ell \ell}$ while in $y$-axis indicates the statistical error in each bin} 
   \label{fig:pTll}
\end{figure}

The study of finding the contribution of \texttt{dim-8} operators transformed from the couplings of \texttt{dim-6} operators for the  $2\ell2\nu$ final state process is carried out by using the $p_T^{\ell\ell}$ distribution  for each anomalous couplings with the implementation of all the cuts in Table~\ref{tab:cutTable}  are applied up to the cut on that variable. The contribution of aTGCs is presented by using an effective vertex function approach in Ref.~\cite{Degrande_2014}. Because of the aTGCs sensitivity have a potential to each out the high$-p_T$ sector, the $p_T^{\ell\ell} > 110$ GeV bins of the Fig.~\ref{fig:pTll}  are taken into account in the analysis besides that the corresponding number of events are summarized in Table~\ref{tab:eventNumbers}.
 
\begin{table}[!htp]
\caption{Obtained event numbers for the background and signal (where all couplings set to zero, apart from one we are working on it) of $p_{T}^{\ell \ell}$ vairable after Cut-10 are given at FCC-hh with $\mathcal{L}_{int}=$ 10 ab$^{-1}$ \label{tab:eventNumbers} }
\centering{}%
\resizebox{\textwidth}{!}{

\begin{tabular}{lcccl } 
\hline 
Couplings & Signal & Background   & Total \tabularnewline
\hline 
$C_{\widetilde{B}W} / \Lambda^{4}=5.0$ TeV$^{-4}$     & 637,303 & 450,699 & 1088,002  \tabularnewline
$C_{BW} / \Lambda^{4}=5.0$ TeV$^{-4}$      		    & 584,240 & 450,699 & 1034,939  \tabularnewline
$C_{BB} / \Lambda^{4}=5.0$ TeV$^{-4}$   		    & 597,036 & 450,699 & 1047,735  \tabularnewline
$C_{WW} / \Lambda^{4}=5.0$ TeV$^{-4}$    		    & 587,309 & 450,699 & 1038,008  \tabularnewline
\hline 
\end{tabular}
}
\end{table}

We applied $\chi^2$ test with and without including systematic error for finding 95\% C.L. bounds on the each couplings. This $\chi^2$ function is defined as follows

\begin{equation}
\label{eqn:chi2def}
\chi^{2} =\sum_i^{n_{bins}}\left(\frac{N_{i}^{NP}-N_{i}^{B}}{N_{i}^{B}\Delta_i}\right)^{2}
\end{equation}
where $N_{i}^{NP}$ is for the total number of events in the existence of effective couplings, $N_{i}^{B}$ is total number of events of the corresponding SM backgrounds in $i$th bin of the $p_T^{\ell\ell}$ distribution, $\Delta_i=\sqrt{\delta_{sys}^2+ 1 / N_i^B}$ is including the systematic ($\delta_{sys}$) and statistical errors in each bin. 

The presence of aNTGCs will give rise to enhance the yield of events at $p_T^{\ell\ell}$ distributions of the $\ell\ell\nu\nu$ final state of the $ZZ$ process. The bounds on possible contributions from aNTGCs are obtained by using this distribution. To examine the viability of the EFT approach, one needs to require the lowest value of the coefficients to set the operator scale $\Lambda$ beyond the reach of the kinematical range of the distributions in order for the EFT approach not to break down. The coefficients of the \texttt{dim-8} operators could be related to the new physics characteristic scale $\Lambda$~\cite{SENOL2018365}. An upper limit can be enhance the new physics scale $\Lambda$ using the fact that the fundamental theory is strongly coupled. We find $\Lambda<\sqrt{4\pi v\sqrt{s}}\sim17.5$ TeV under the assumption of the couplings $C=O(1)$. This upper bound is not violated in this analysis as we have $m_{\ell\ell}<1.2$ TeV for the kinematic range of invariant mass distributions.

\begin{figure} [hbt]
 \centering
   \includegraphics[width=1\textwidth]{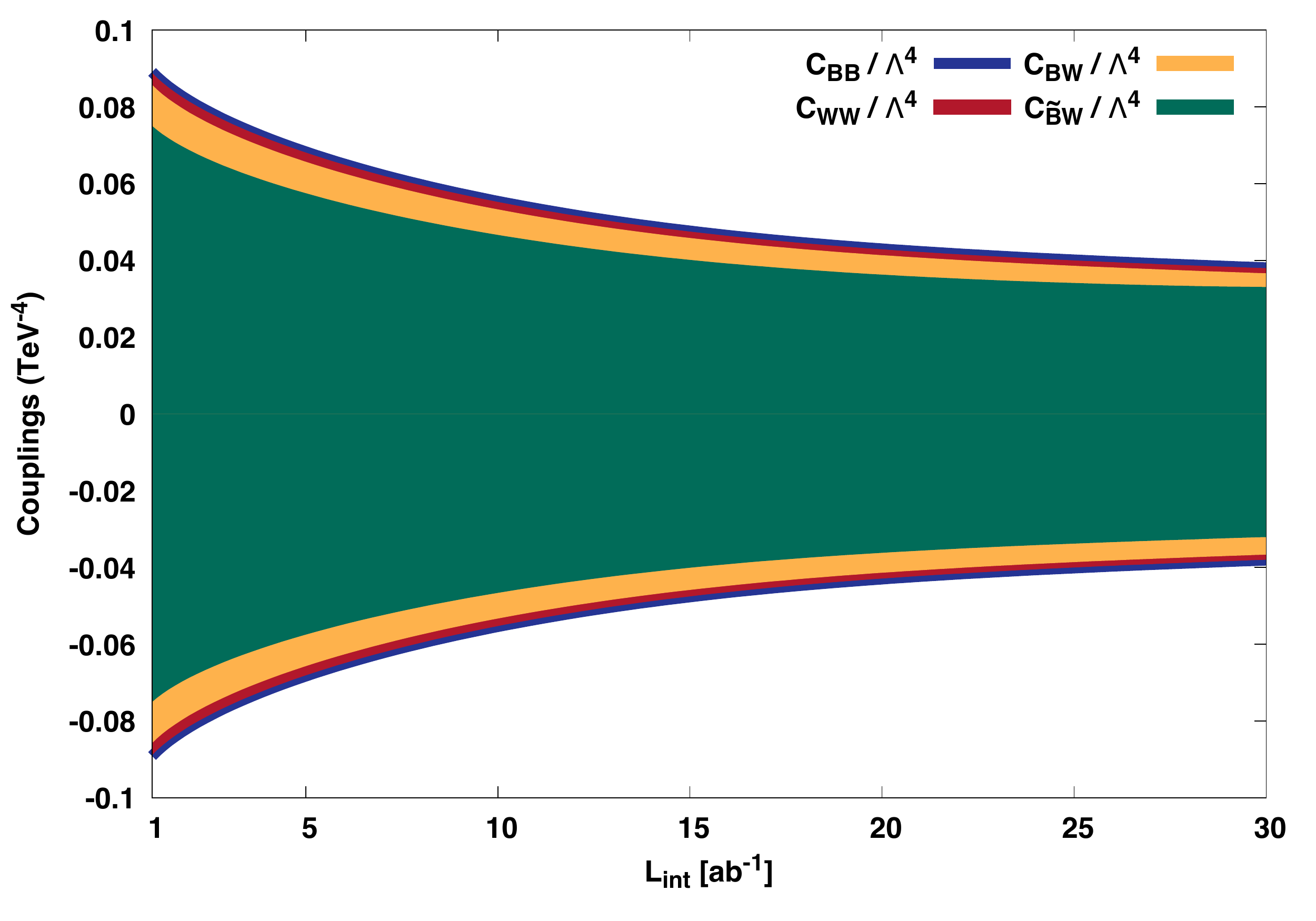}
  \caption{Estimation of extracted limits on aNTG couplings at $95\%$ C.L. (without systematic error) as a function of integrated luminosity in unit of [ab$^{-1}$]. the computation is made with changing one coupling at a time from its SM value} 
   \label{fig:sensitivityLimit}
\end{figure}

Analyzing of $ZZ$ production with $2\ell2\nu$ in the final state, the number of signal events and one-parameter $\chi^2$ results for each couplings varied with integrated luminosity from 1 ab$^{-1}$ to 30 ab$^{-1}$. In the analysis, only one coupling at a time is varied from its SM value. 
The estimated results from $\chi^2$ analysis of the couplings describing aTGC interactions of neutral gauge bosons. The coefficients of the operators denoted as $C_{\widetilde{B}W}  / \Lambda^{4}$,  $C_{WW} / \Lambda^{4}$, $C_{BW} / \Lambda^{4}$ and  $C_{BB}  /   \Lambda^{4}$ are given in Fig.~\ref{fig:sensitivityLimit}. 

The obtained of one-dimensional 95\% \textit{C.L.} bounds at $L_{int}$ = 10 ab$^{-1}$  with and without including  the effects of systematic errors on the limits are summarized in Table~\ref{tab:obtainedLimits} assuming that any excess in signal over background expected solely to contribution of $C_{\widetilde{B}W} / \Lambda^{4}$, $C_{WW}  / \Lambda^{4}$, $C_{BW} /  \Lambda^{4}$ or $C_{BB}  /  \Lambda^{4}$ couplings.

\vspace*{0.3cm}

\begin{table}[h!]
\centering
\caption{Estimation of new bounds values of aNTG couplings with and without a systematic error at $L_{int}$ = 10 ab$^{-1}$  at 95\% C.L. All parameters other than the one under study are set to zero  for each single anomalous coupling }

\resizebox{1\textwidth}{!}{
\begin{tabular}{ lccc}
 \hline
 Couplings & \multicolumn{3}{c}{Limits at 95\% C.L.} \\
    $ (TeV^{-4})$  & $\delta_{sys}= 0\%$ & $\delta_{sys}= 1\%$ & $\delta_{sys}= 3\%$   \\
\hline
$C_{\widetilde{B}W}  / \Lambda^{4}$   & $[-\,0.042, \,\, +\,0.042]$  & $[-\,0.110, \,\, +\,0.110]$  & $[-\,0.189, \,\, +\,0.189]$    \tabularnewline
$C_{WW}  / \Lambda^{4}$                   & $[-\,0.050,\,\, +\,0.050]$         & $[-\,0.131, \,\, +\,0.131]$ & $[-\,0.225, \,\, +\,0.225]$  \tabularnewline
$C_{BW} /   \Lambda^{4}$                   & $[-\,0.050,\,\, +\,0.050]$         & $[-\,0.131, \,\, +\,0.131]$ & $[-\,0.225, \,\, +\,0.225]$  \tabularnewline
$C_{BB}  /   \Lambda^{4}$                   & $[-\,0.048, \,\, +\,0.048]$  & $[-\,0.126, \,\, +\,0.126]$ & $[-\,0.217, \,\, +\,0.217]$   \tabularnewline
\hline 
\end{tabular} 
}
\label{tab:obtainedLimits}
\end{table}


\section{Conclusion} \label{sec:conclusion}

A phenomenological cut based study for searching the bounds of \texttt{dim-8} aNTG  CP-conserving $C_{\widetilde{B}W} / \Lambda^{4}$ and CP-violating $C_{WW}  / \Lambda^{4}$, $C_{BW} /  \Lambda^{4}$ and $C_{BB} / \Lambda^{4}$  \texttt{dim-8} aNTG couplings via  $ZZ \rightarrow \ell \ell \nu \nu$ (where $\ell$ =  $e$ or $\mu$) production at the FCC-hh is summarized in this paper.
When we compare the obtained bounds of \texttt{dim-8} aNTG couplings at 95\% C.L. with latest results of LHC~\cite{Sirunyan:2020pub}, the sensitivity of each aNTG couplings  is improved. 
The obtained results without systematic error, 53\% ($C_{\widetilde{B}W} / \Lambda^{4}$ ), 76\% ($C_{WW}  / \Lambda^{4}$), 81\%  ( $C_{BW} /  \Lambda^{4}$) and 52\% ($C_{BB} / \Lambda^{4}$), are better than the current phenomenological study~\cite{Yilmaz_2020}  which is done for the process $pp\rightarrow ZZ \rightarrow 4\ell$ at \verb"FCC-hh" with an integrated luminosity $L_{int} = 10$ fb$^{-1}$.

Even with $3\%$ systematic errors, the obtained limits for FCC-hh are comparable or better than current LHC results. The obtained limits on aNTG couplings from this study would be to the advantage of the high luminosity when the systematic uncertainties are well reduced below $1\%$.
These are current upper bounds on aNTG couplings.

\section*{References}
\bibliography{pp2zz_llvv2021}

\end{document}